\documentclass{fundam}

\usepackage{url} 
\usepackage[ruled,lined]{algorithm2e}

\usepackage{lineno,hyperref}
\modulolinenumbers[5]

\usepackage{graphicx}
\usepackage{amssymb,amsmath}
\usepackage{algpseudocode}
\usepackage{enumitem}
\newcommand{\Real}{\mathbb{R}}

\newenvironment{claim_proof}{\paragraph{Proof:}}{\hfill$\lozenge$}

\mathchardef\mhyph="2D

\begin{document}

\setcounter{page}{1001}
\issue{XXI~(2001)}

\title{Recognizing Visibility Graphs of Triangulated Irregular Networks}

\author{Hossein Boomari\\
Department of Mathematical Science \\
Sharif University of Technology \\ 
Tehran, Iran\\
h.boomari1@student.sharif.ir
\and Mojtaba Ostovari\\
Department of Mathematical Science \\
Sharif University of Technology \\
Tehran, Iran \\
m.ostovari@student.sharif.ir
\and Alireza Zarei \\
Department of Mathematical Science \\
Sharif University of Technology \\
Tehran, Iran \\
zarei@sharif.ir} 

\maketitle

\runninghead{H. Boomari, Mojtaba. Ostovari, A. Zarei}{Recognizing Visibility Graphs of TINs}

\address{Sharif University of Technology, Azadi Street, Tehran, Iran}

\begin{abstract}

A \textit{Triangulated Irregular Network} (\textit{TIN}) is a data structure that is usually used for representing and storing monotone geographic surfaces, approximately. In this representation, the surface is approximated by a set of triangular faces whose projection on the $XY$-plane is a triangulation.
The visibility graph of a \textit{TIN} is a graph whose vertices correspond to the vertices of the \textit{TIN} and there is an edge between two vertices if their corresponding vertices on \textit{TIN} see each other, i.e. the segment that connects these vertices completely lies above the \textit{TIN}.


Computing the visibility graph of a \textit{TIN} and its properties have been considered thoroughly in the literature. In this paper, we consider this problem in reverse: Given a graph $G$, is there a \textit{TIN} with the same visibility graph as $G$? We show that this problem is $\exists \Real\mhyph Complete$.


\end{abstract}



\section{Introduction}
\begin{figure}[ht]
\centerline{\includegraphics[scale=0.25]{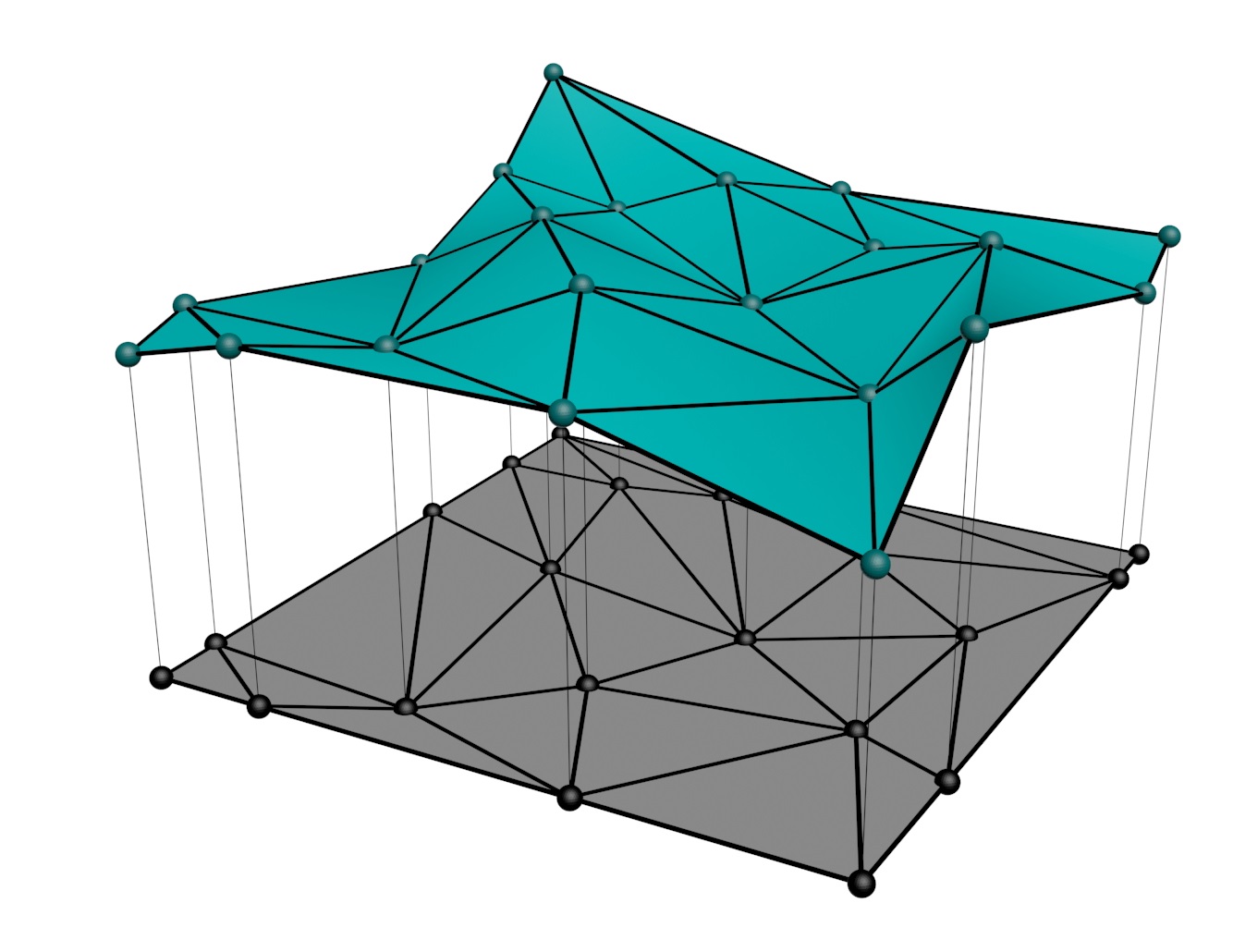}}
\caption{A Triangulated Irregular Network (\textit{TIN}) and its planar projection.}\label{fig:1}
\end{figure}

A monotone surface is a surface with at most one intersection point with any vertical line. Such a surface can be represented as a function $S:\Real^2\rightarrow\Real$. As a linear estimation, such surfaces are represented by \textit{Triangulated Irregular Network}s (\textit{TIN}s) which are composed of a set of triangular faces. Projection of a \textit{TIN} on the $XY$-plane is therefore a triangulation in the plane (See Fig.~\ref{fig:1}).

The visibility relations between vertices of a \textit{TIN} can be represented by a graph called its \textit{visibility graph}.
In this graph, each vertex corresponds to a \textit{TIN}'s vertex and there is an edge between a pair of vertices, if and only if the segment between their corresponding vertices on the \textit{TIN} lies completely above the \textit{TIN}.
The visibility graphs have applications in many computational geometry problems such as motion planning and ray tracing, computer graphics, robotics and other fields which consider the geometry of surfaces like \textit{GIS}\footnote{Geographic information system} and geology~\cite{visgeo}. Therefore, the problem of computing the visibility graph has been considered thoroughly and there are several polynomial time algorithms for computing these graphs for \textit{TIN}s~\cite{visgraphic,viscomp}.
The visibility graph is also defined for some other classes of geometric shapes like polygons, monotone curves and points in the plane.

Considering the problem of computing the visibility graph in reverse is another interesting theoretical problem in computational geometry~\cite{PVG,ghosh3}. In this problem we are asked to determine whether there is a realization in the form of a specific geometric object for a given visibility graph.
Restricting this problem for \textit{TIN}s, the goal is to determine whether there exists a \textit{TIN} whose visibility graph is isomorphic to a given graph. This problem is known as \textit{recognizing visibility graphs} and mainly is focused on defining necessary and sufficient conditions on a graph to be a visibility graph. There are different variations of this recognition problem with different constraints on the requested \textit{TIN}, the description of the input of the problem, or the class of the given input graph. For example, in two variations of this problem, the triangulation graph of the vertices of the \textit{TIN} may or may not be given in the input. If the answer to the recognizing problem for a given graph with the specified constraints is "yes", the next question is to build such a \textit{TIN} which is known as \textit{reconstructing visibility graph} problem. These problems have also been defined for other geometric shapes like simple polygons in which the target geometric object is a simple polygon in the plane. Beside the variety of visibility graphs with respect to the classes of geometric shapes, there is another problem based on pseudo-visibility. There are some results for the realization and characterization of pseudo-visibility graphs~\cite{pseudo_visibility}.

Despite much progress in computing the visibility graphs, the computational complexity of the recognition and reconstruction problems are still open for almost all general shapes, including simple polygons, terrains, and \textit{TIN}s~\cite{terrain, matroid}. It is only known that these problems belong to \textit{PSPACE}, and more precisely, to the \textit{Existential Theory of The Reals} ($\exists \Real$) complexity class.

Existential theory of the reals ($\exists \Real$) is the complexity class that was implicitly introduced in 1989~\cite{npnpc} and explicitly defined by Shor in 1991~\cite{shor1991}. Schaefer shows that it is the complexity class of the problems which can be reduced, in polynomial time, to the problem of deciding, whether there is a solution for a Boolean formula $\phi:\{True,False\}^n\rightarrow \{True,False\}$ in propositional logic, in the form
$\phi(F_1(X_1,X_2,...,X_N) , F_2(X_1,X_2,...,X_N), ..., F_n(X_1,X_2,...,X_N))$,
where each $F_i:\Real^N \rightarrow \{True, False\} $ consists of a polynomial function $G_i:\Real^N\rightarrow \Real$ on some real variables, compared to $0$ with one of the comparison operators in $\{<,\leq,=,>,\geq\}$ (for example $G_i(X_1,X_2)=X_1^3X_2^2-5X_1X_2^3+8$ and $F_i(X_1,X_2) \equiv G_i(X_1,X_2)<0$)~\cite{schaefer}. Equivalently, it is the complexity class of the problems which are polynomial-time reducible to the problem of deciding the emptiness of a semi-algebraic set~\cite{exist}. Clearly, satisfiability of a quantifier-free boolean formula belongs to $\exists \Real$. Therefore, $\exists \Real$ includes all $NP$ problems. In addition, $\exists \Real$ belongs to $PSPACE$~\cite{exist} and we have $NP \subseteq \exists \Real \subseteq PSPACE$. 
Although this class is inherently an Algebraic complexity class, it has drawn the attention from geometers because some of the main arguments in its literature rely on geometric properties. Another reason is that some geometric problems are \textit{complete} (with respect to polynomial-time reductions) for this class. For example Recognizing \textit{LineArrangement}, \textit{Stretchability}, \textit{Simple Order Type}, \textit{Intersection Graph of Segments}, and \textit{Intersection Graph of Unit Disks} in the plane are complete for $\exists \Real$ or simply $\exists \Real\mhyph  Complete$~\cite{ERP}. As the most related result to this paper, in 2017 Cardinal \textit{et al.} showed that recognizing visibility graph of a point set is $\exists \Real\mhyph Complete$~\cite{PVG}. The visibility graph of a set of 3D-points is a graph with these points as its vertices and there is an edge between two vertices if their connecting segment does not pass through any other point.

Because of their application in our proofs, we discuss Recognizing \textit{LineArrangement}  and \textit{Stretchability} problems with more details in Section~\ref{sec:B}. In Section~\ref{sec:C}, we consider the recognition problem of \textit{TIN}s and prove that this problem is also $\exists \Real\mhyph Complete$. In this problem, we are given a pair of a graph and a triangulation which are respectively the visibility graph and the triangulation of a target \textit{TIN} on the plane. Then, the question is whether there is a 3D \textit{TIN} realization for this triangulation whose visibility graph is the same as the given graph. The formal description of this problem is:

Recognition of visibility graphs of polygonal terrains:

INPUT: A pair $\langle G,T \rangle$, where $G$ and $T$ are two graphs on the same vertex set $V$, and $T$ is a planar subgraph of $G$ with a planar embedding whose all faces, except possibly its external face $f$, are triangular ($f$ is not mentioned in the input explicitly). 

QUESTION: Does there exist a \textit{TIN} with vertex set $V$, whose projection on the plane is an embedding of $T$ with outer face $f$, and whose visibility graph is $G$?



\section{Preliminaries and Definitions}
\label{sec:B}
\subsection{LineArrangement and Stretchability}
Combinatorial description of a geometric shape in the plane is an interesting problem in both theory and applications of computational geometry. \textit{LineArrangement} is an example of such descriptions for a set of lines in the plane. \textit{LineArrangement} describes the leftmost vertical order of the lines in the plane (initial order) and the left to right order of intersections of each line by the other lines (this ordering, for a line $L_i$, is denoted by $Seq(L_i)$). This arrangement is well defined for an arbitrary set of lines in the plane, including when the lines are in \textit{general position}\footnote{An arrangement which has no pair of parallel lines and no triple of lines intersecting at the same point} as we will assumed in this paper. Recognizing \textit{LineArrangement} for a given line arrangement instance, is the problem of deciding whether there is a set of lines in the plane with the same arrangement as the input. This problem is $\exists \Real\mhyph Complete$~\cite{ERP}.

Similar to the definition of the monotone surfaces in the Introduction section, we use the term \textit{monotone curve} for a curve in the plane which intersects any vertical line exactly once. A set of pseudo-lines in the plane is a set of monotone curves such that each pair intersect exactly once.  Based on this property, \textit{PseudoLineArrangement} (and its recognition problem), is defined analogously. 
In contrast to \textit{LineArragement}, recognizing a \textit{PseudoLineArrangement} can be solved efficiently in polynomial time
~\cite{allow}. This difference introduces another problem called \textit{Stretchability}, which is stated as: ``Is it possible to stretch a pseudo-line realization of a \textit{PseudoLineArrangement} and make them a set of lines without changing their arrangement?". \textit{PseudoLineArrangement} belongs to $P$ and recognizing \textit{LineArrangement} is $\exists \Real\mhyph Complete$; proving that the stretchability problem is $\exists \Real\mhyph Complete$.

\subsection{From PseudoLineArrangement to Triangulation}
\begin{figure}[ht]
\centerline{\includegraphics[scale=0.60]{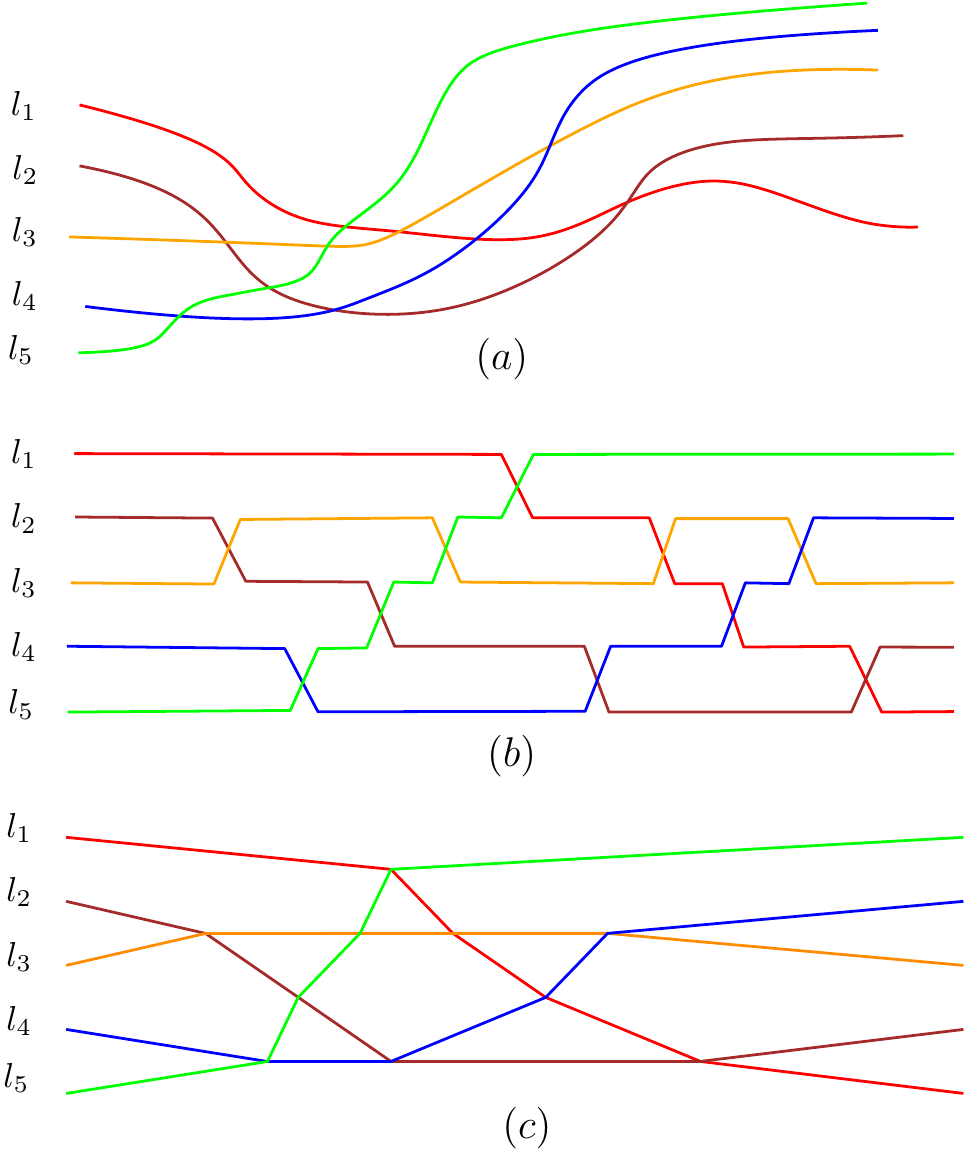}}
\caption{The reconstruction algorithm for \textit{PseodoLineArrangement}. Algorithm~\ref{alg:1} in Appendix section contains a pseudo code implementation of this algorithm.\label{fig:2} }
\end{figure}

\begin{figure}
\centerline{\includegraphics[scale=0.50]{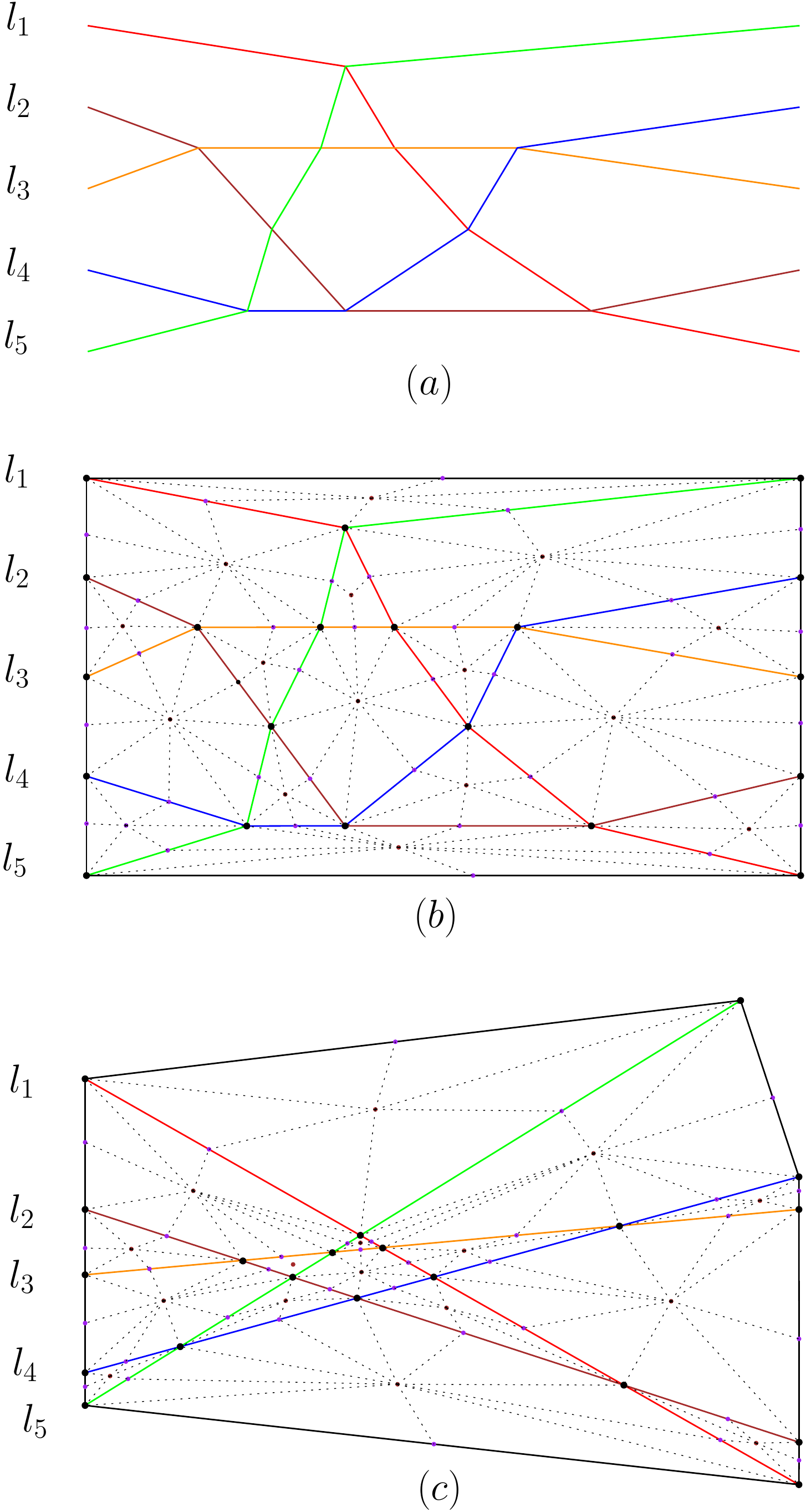}}
\caption{(a) The \textit{PseudoLineArrangement} realization for an instance with initial order $\langle L_1,L_2,L_3,L_4,L_5 \rangle $, $Seq(L_1)=\langle L_5,L_3,L_4,L_2 \rangle $, $Seq(L_2)=\langle  L_3,L_5,L_4,L_1\rangle $, $Seq(L_3)=\langle L_2,L_5,L_1,L_4 \rangle $, $Seq(L_4)=\langle  L_5,L_2,L_1,L_3 \rangle $ and $Seq(L_5)=\langle  L_4,L_2,L_3,L_1 \rangle $, (b) The corresponding triangulation for the \textit{PseudoLineArrangement} realization as an instance of \textit{StretchableTriangulation} (c) A realization for the instance of  \textit{StretchableTriangulation} which is also a realization for the initial \textit{LineArrangement} instance.\label{fig:3}}
\end{figure}

Similar to recognizing \textit{PseudoLineArrangement}, its reconstruction\footnote{Computing a set of planar pseudo-lines with the given arrangement} problem also belongs to $P$ and can be computed efficiently~\cite{allow} (See Fig.~\ref{fig:2}-a and Fig.~\ref{fig:2}-b).  According to the given implementation of the reconstruction algorithm depicted in Algorithm~\ref{alg:1} in Appendix section, which is a variation of the algorithm presented in~\cite{allow} with a few modifications, we can obtain a special reconstruction of the pseudo-lines in which each pseudo-line is composed of a sequence of segments and the joint points of these segments correspond to the intersection points of the arrangement, except the first and the last point of each pseudo-line which are arbitrary points before leftmost and after the rightmost points on that pseudo-line. We call these leftmost and rightmost points of each pseudo-line the first and the last end-point of that pseudo-line, respectively. More precisely, each segment connects an intersection between pseudo-lines to the next one, except the first and the last end-points of each pseudo-line which connects the first (resp. the last) point on a pseudo-line to the first (resp. last) intersection point of that pseudo-line (See Fig.~\ref{fig:2}-c). This special reconstruction of a \textit{PseudoLineArrangement} partitions the plane into convex regions. 
The regions are convex because the algorithm draws any of the four segments connected to a break-point $p$ inside a different quadrant around $p$ which forces all the angles around $p$ to be convex. 
For such a subdivision, we associate a triangulation as follows: We put a new vertex (splitting vertices) on each segment of each pseudo-line, and split each edge into two edges with slightly different slopes (See Fig.~\ref{fig:3}-b). We replace the new segments which are made from the original segments and these new segments belong to the same pseudo-line. In addition, we add a vertex (middle vertices) inside each convex region. Then, each middle vertex is connected to all vertices on the boundary of its region (See Fig.~\ref{fig:3}-b). The resulting subdivision is a triangulation. 

For an instance $\mathcal{A}$ of \textit{PseudoLineArrangement}, this special reconstruction is denoted by $\mathcal{S}_\mathcal{A}$; the corresponding triangulation is denoted by $\mathcal{T}_\mathcal{A}$; $\mathcal{S}_\mathcal{A}(P)$ and $\mathcal{T}_\mathcal{A}(P)$ are respectively the sequences of segments of a pseudo-line $P$ in $\mathcal{S}_\mathcal{A}$ and $\mathcal{T}_\mathcal{A}$; the first end-point of $P$ in $\mathcal{S}_\mathcal{A}$ and $\mathcal{T}_\mathcal{A}$ are respectively denoted by $\mathcal{S}_\mathcal{A}^{first}(P)$ and $\mathcal{T}_\mathcal{A}^{first}(P)$, and the last ones are respectively denoted by $\mathcal{S}_\mathcal{A}^{last}(P)$ and $\mathcal{T}_\mathcal{A}^{last}(P)$. 
The corresponding graph of $\mathcal{T_\mathcal{A}}$, whose vertices and edges respectively correspond to the vertices and edges of the triangulation, is called the triangulation graph of $\mathcal{T}_\mathcal{A}$.

\begin{lemma}
\label{lem:1}
For each pair of adjacent triangles $t_1$ and $t_2$ in $\mathcal{T}_\mathcal{A}$ with vertex sets $\{a,b,c\}$ and $\{a,b,d\}$ in  $\mathcal{T}_\mathcal{A}$, the edge $(c, d)$ does not exist in $\mathcal{T}_\mathcal{A}$ ($\mathcal{T}_\mathcal{A}$ has no $K_4$).
\end{lemma}
\begin{proof}
The common edge of $t_1$ and $t_2$ is either a part of a pseudo-line, or a segment that connects a vertex of a pseudo-line to a newly added vertex inside a region (the middle vertices) of the \textit{PseudoLineArrangement} realization $\mathcal{S}_\mathcal{A}$. In the former case, the non-common vertices $c$ and $d$ are necessarily two newly added vertices inside different regions and there is no edge between them. In the latter case, the  vertices $c$ and $d$ are two non-adjacent vertices on pseudo-lines (may be on a single pseudo-line). According to the process of building $\mathcal{T}_\mathcal{A}$, there is no triangle whose all vertices are on the pseudo-lines. Therefore, in both cases $c$ and $d$ are not adjacent in $\mathcal{T}_\mathcal{A}$.
\end{proof}

On the other hand, we can obtain a realization for an instance $\mathcal{A}$ of \textit{PseudoLineArrangement} from a realization of its corresponding $\mathcal{T}_\mathcal{A}$. This is done by removing the added middle and splitting vertices and their adjacent edges from $\mathcal{T}_\mathcal{A}$ and adding segments to connect each pair of vertices separated by a splitting vertex. As the middle vertices do not belong to pseudo-lines and the splitting vertices are the vertices with odd indices in the sequence of the vertices of each pseudo-line (staring the indices from $0$) in $\mathcal{T_A}$, these vertices and edges can be distinguished efficiently. 

Therefore, \textit{Stretchability} can be reduced in polynomial time to the problem of whether there is a triangulation $\mathcal{T}_\mathcal{A}$ in the plane in which each pseudo-line in $\mathcal{T}_\mathcal{A}$ lies along a single line (see Figure~\ref{fig:3}-c). We call this problem as \textit{StretchableTriangulation}. Here is the definition of \textit{StretchableTriangulation}:

Recognition of \textit{StretchableTriangulation}:

INPUT: A triangulation graph $\mathcal{T}_\mathcal{A}$ without $K_4$ and a set of paths on this graph which are called pseudo-lines.

QUESTION: Does there exist a realization for $\mathcal{T}_\mathcal{A}$, in which each pseudo-line lies along a single line?

 We can summarize the above discussion as the following theorem.

\begin{theorem}
\label{thm:2}
\textit{Stretchability} can be reduced to \textit{StretchableTriangulation} in polynomial time and therefore, \textit{StretchableTriangulation} is $\exists \Real\mhyph Complete$.
\end{theorem}

Note that this theorem is true for triangulation graphs with $K_4$ as well, but we concentrated on having no $K_4$ to use this feature in our next theorems.

\section{From StretchableTriangulation to Recognizing Visibility Graphs of \textit{TIN}s}
\label{sec:C}
We prove that deciding whether a triangulation $\mathcal{T}_\mathcal{A}$ is stretchable, can be reduced to an instance of the problem of recognizing visibility graphs of  \textit{TIN}s. For this purpose, for an instance ${\mathcal{T}_\mathcal{A}}$ of \textit{StretchableTriangulation}, we build an instance $\langle G,T \rangle $ of recognizing visibility graphs of \textit{TIN}s problem where $\mathcal{T}_\mathcal{A}$ is stretchable if and only if $\langle G,T \rangle $ is realizable. Remember that in $\langle G,T \rangle $ instance, $G$ is the visibility graph of the vertices of the target \textit{TIN} and $T$ is the corresponding triangulation of the planar projection of this \textit{TIN}. To construct the $\langle G,T \rangle $ instance, $T$ is initially set to be the triangulation $\mathcal{T}_\mathcal{A}$ and $G$ is initialized by exactly the vertex set and edges of this triangulation. Name this initial instance as $I_{\mathcal{T}_\mathcal{A}}$. For a realizable instance of $\mathcal{T}_\mathcal{A}$, the $I_{\mathcal{T}_\mathcal{A}}$ instance of the recognition problem has at least one realization which is exactly the realization of $\mathcal{T}_\mathcal{A}$ in the plane (which is a flat \textit{TIN}). The next theorem states that all realizations of $I_{\mathcal{T}_\mathcal{A}}$ are concave \textit{TIN}s, i.e. the angle between each pair of adjacent faces is concave, seeing the \textit{TIN} from above.

\begin{figure}[ht]
\centerline{\includegraphics[scale=0.15]{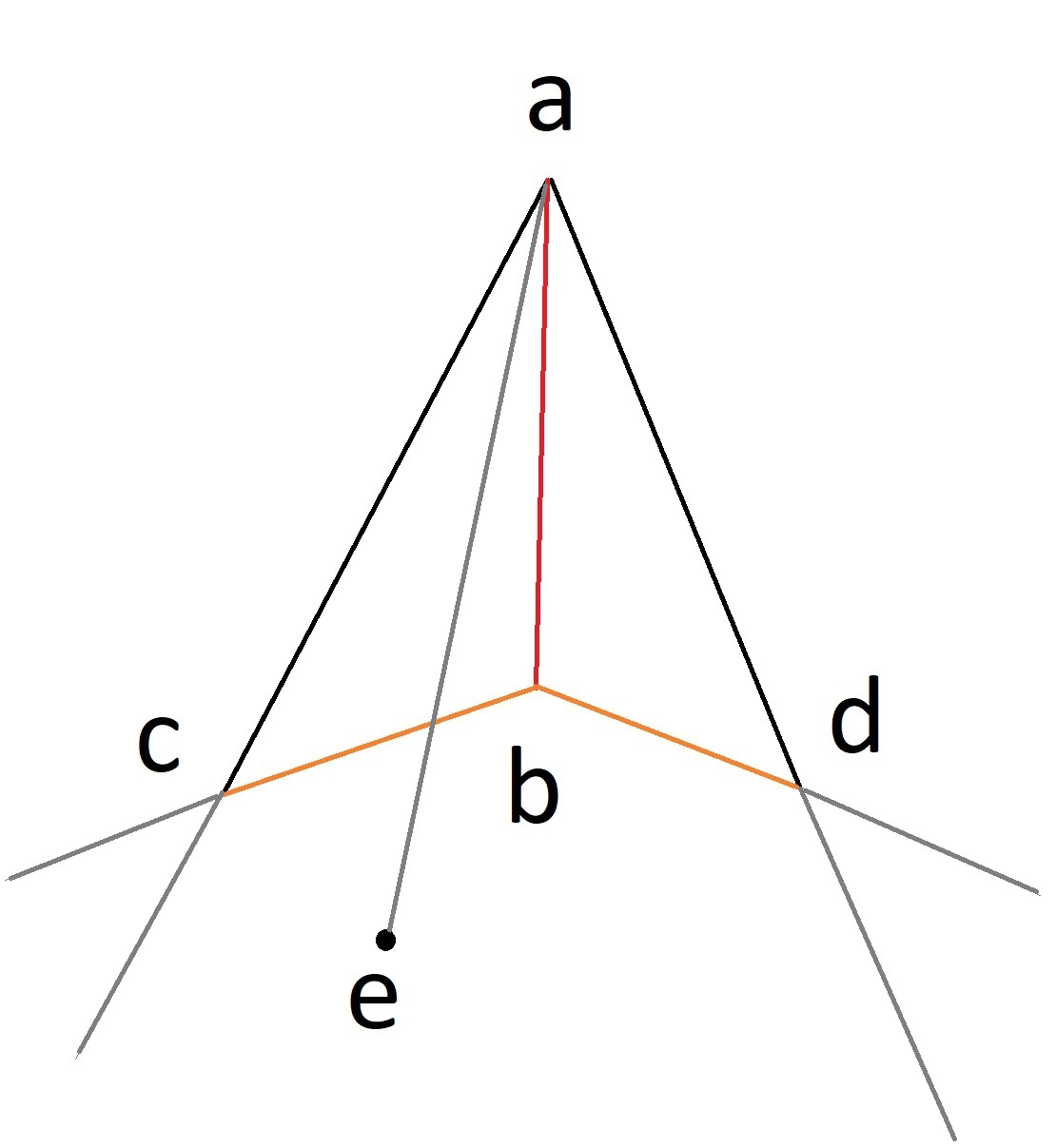}}
\caption{There can be no adjacent faces with a convex angle in any realization of $I_{\mathcal{T}_\mathcal{A}}$. \label{fig:thm} }
\end{figure}

\begin{lemma}
\label{thm:3}
Any realization of the above $I_{\mathcal{T}_\mathcal{A}}$ instance of recognizing visibility graphs of \textit{TIN}s 
is concave.
\end{lemma}
\begin{proof}
For the sake of a contradiction, assume that there is a non-concave realization $\mathcal{R}$ for an instance $I_{\mathcal{T}_\mathcal{A}}$ of \textit{RecognzingVisibilityGraphsOfTINS}. There must be at least two adjacent faces $A$ and $B$ in $\mathcal{R}$ whose interior angle is convex. Let $\{a, b, c\}$ and $\{a, b, d\}$ be respectively the vertices of faces $A$ and $B$ (see Fig.~\ref{fig:thm}). For a vertex $v$ on the realized TIN, denote $v_T$ as its corresponding vertex in $T$. Lemma~\ref{lem:1} states that $c_T$ and $d_T$ are not adjacent in $T$,  and hence should not be visible in $G$. Therefore, the visibility of the pair $(c, d)$ must be blocked by some part of the \textit{TIN} that lies above the plane through vertices $b$, $c$ and $d$. Because of the monotonicity of the \textit{TIN}, there must be at least one vertex, like $e$, in this part of the \textit{TIN} which is visible from $a$ such that the edge $e_Ta_T$ does not exist in $T$. The reason of non-adjacency of $a_T$ and $e_T$ in $T$ is that otherwise either $ae$ must break the monotonicity of the \textit{TIN} or $a_Te_T$ intersect $c_Tb_T$ or $b_Td_T$ which is in contradiction with the existence of the faces $A$ and $B$. Consequently, $a$ and $e$ must have a visibility edge in $T$ which is impossible according to Lemma~\ref{lem:1}.
\end{proof}

\begin{figure}[ht]
\centerline{\includegraphics[scale=0.5]{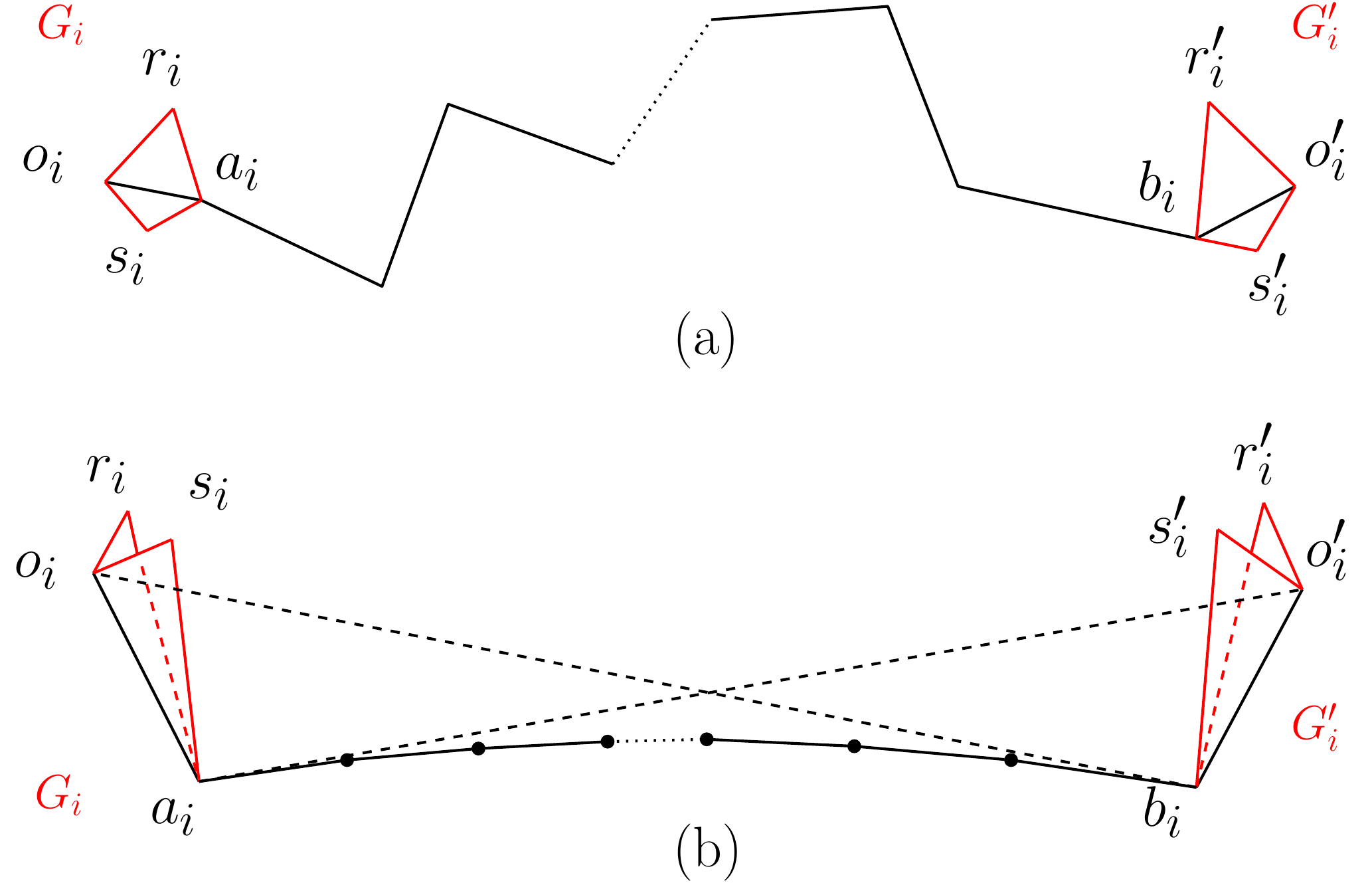}}
\caption{Layout of alignment gadgets from a) top view b) side view. \label{fig:4} }
\end{figure}

\begin{figure}[ht]
\centerline{\includegraphics[scale=0.5]{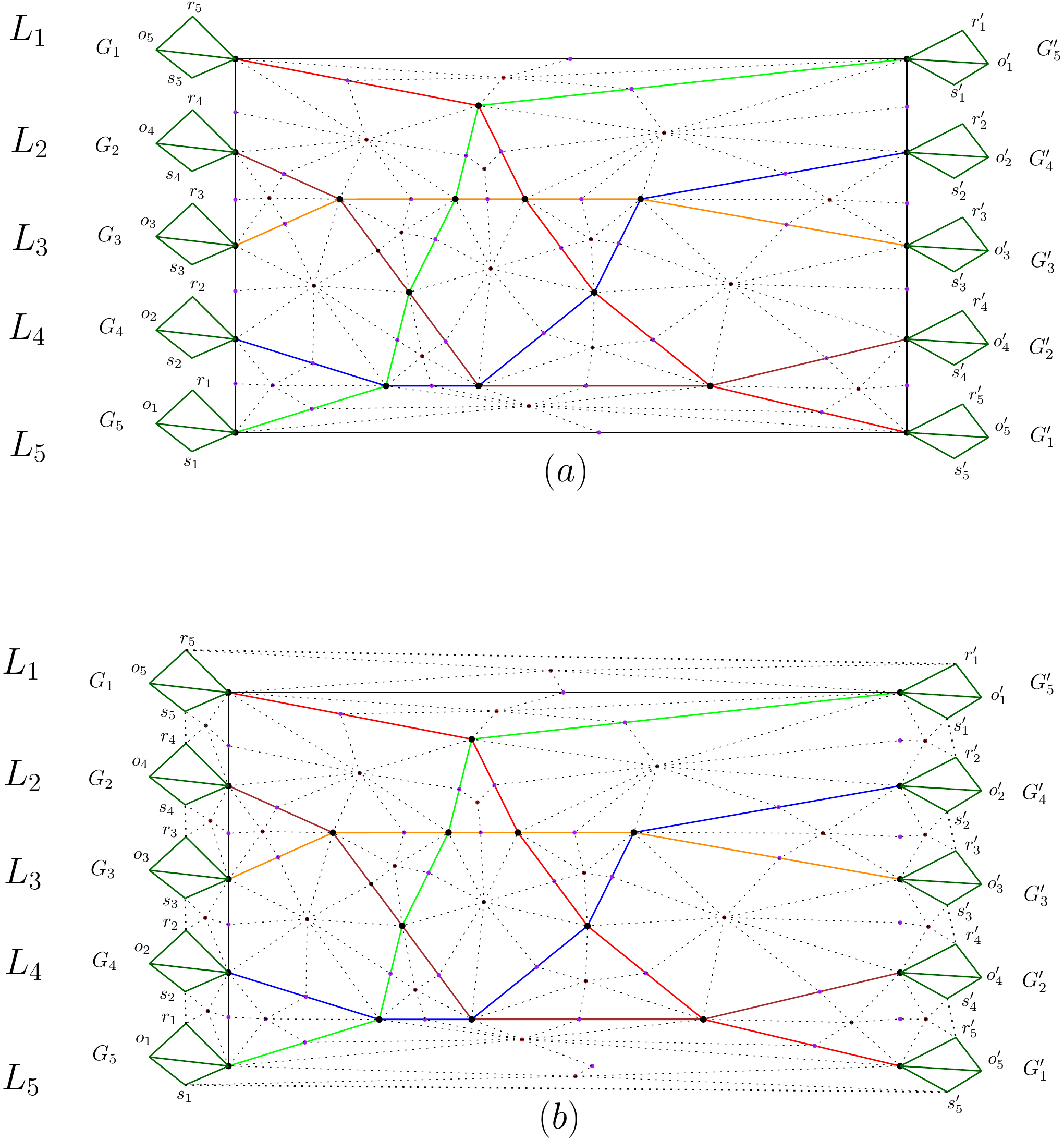}}
\caption{Reducing $\mathcal{T}_\mathcal{A}$ to $\langle G,T \rangle$ (a) adding alignment gadgets (b) refining the triangulation.\label{fig:5} }
\end{figure}

To complete the reduction (build $\langle G,T \rangle$ from $I_{\mathcal{T}_\mathcal{A}}$), we need a mechanism to enforce that in all $\mathcal{T}_\mathcal{A}$ realizations, each pseudo-line $P_i$ lies along a single line. 
For this purpose, we attach gadgets, called alignment gadgets, to both corresponding vertices of $\mathcal{T}_\mathcal{A}^{first}(P_i)$ and $\mathcal{T}_\mathcal{A}^{last}(P_i)$ in $G$ and $T$. The structures of this pair of gadgets for any pseudo-line $P_i$, restrict $P_i$, in any realization of $\langle G,T \rangle $, to lie in an arbitrarily narrow convex volume. As shown in Fig.~\ref{fig:4}, the start and end gadgets of $P_i$ that are connected to the first and last vertices $a_i$ and $b_i$ of a pseudo-line $P_i$ are respectively $\{s_i,r_i,o_i\}$ and $\{s^\prime_i,r^\prime_i,o^\prime_i\}$ vertices, where all three vertices in each gadget are connected to their corresponding end-point in triangulation $T$. Moreover, to be a valid triangulation, there are edges from $o_i$ (resp. $o^\prime_i$) to both vertices $s_i$ and $r_i$  (resp. $s^\prime_i$ and $r^\prime_i$) in $T$. As shown in Fig.~\ref{fig:5}.a, this triangulation still needs some extensions to be a polished complete triangulation. This is done by adding some extra vertices and edges as shown in Fig.~\ref{fig:5}.b. This refinement is formally obtained by applying the following changes (assume that $\langle \{s_1,r_1,o_1\},...,\{s_n,r_n,o_n\},\{s^\prime_1,r^\prime_1,o^\prime_1\},...\{s^\prime_n,r^\prime_n,o^\prime_n\} \rangle$ is the order of the gadgets around the outer boundary of $\mathcal{T}_\mathcal{A}$):
\begin{enumerate}[label=T-\arabic*.,itemindent=*]
\item Add edges $(r_i,s_{i+1})$ and $(s^\prime_i,r^\prime_{i+1})$ for $1 \leq i \leq n-1$ and the two edges $(r_n,r^\prime_1)$ and $(s_1,s^\prime_n)$
\item Add a vertex in each region $(r_i,s_{i+1},a_{i+1},a_i)$ and $(s^\prime_i,r^\prime_{i+1},b_{i+1},b_i)$ for $1 \leq i \leq n-1$ and one vertex in each of the regions $(a_n,r_n,r^\prime_1,b_1)$ and $(s^\prime_n,s_1,a_1,b_4)$ and connect each of these new vertices to the vertices on its region boundary.
\end{enumerate}

Note that $T$ is always a subgraph of $G$ which means that, untill now (before these modifications), all vertices and edges of this refined triangulation exists in $G$, and the next modifications implicitly add these new edges (edges which are added in steps $T-1$ and $T-2$) to $G$. In order to enforce the stretchability of any pseudo-line $P_i$ in any realization of the $\langle G,T \rangle$ instance of the recognition problem, the following visibility edges are added to $G$ to complete the  $\langle G,T \rangle $ instance:

\begin{enumerate}[label=G-\arabic*.,itemindent=*]
\item  For each pair of vertices $p$ and $q$ of $G$ where none is $o_i$ or $o^{\prime}_i$ and $p$ is a vertex of a gadget or a vertex added in step $T\mhyph 2$, the visibility edge $(p,q)$ is added to G.
\item For each pseudo-line $P_i$, the visibility edges from $o_i$ and $o^{\prime}_i$ to all vertices of $G(P_i) \cup \{r_i,s_i,r^{\prime}_i,s^{\prime}_i\} $ as well as the edge $(o_i,o^{\prime}_i)$ are added to $G$.
\end{enumerate}

In Theorem~\ref{thm:4}, we show that this structure forces the stretchability of $P_i$ in any realization of the $\langle G,T \rangle $ instance.

\begin{theorem}
\label{thm:4}
\textit{StretchableTriangulation} is reducible to recognizing visibility graphs of \textit{TIN}s in polynomial time.
\end{theorem}
\begin{proof} 
Clearly, the above construction of the $\langle G,T \rangle $ instance of recognizing visibility graphs of \textit{TIN}s (the visibility graph and the triangulation) can be done in polynomial time. Therefore, we need to prove that there is a realization for the instance $\langle G,T \rangle $ if and only if there is a realization for the corresponding instance of \textit{StretchableTriangulation} $\mathcal{T}_\mathcal{A}$. 

Let $S$ be the set of vertices $\bigcup\limits_{i=1}^{n} \{r_i, s_i, r^{\prime}_i,s^\prime_i\}$ and the vertices added in step $T\mhyph 2$ and $S^\prime$ be the set of all vertices except $o_j$ and $o^\prime _j$ (which are only visible to the vertices of their pseudo-line and its attached allignment gadgets). The visibility graph forces each vertex of $S$ to see all vertices of $S^\prime$. 
Intuitively, in any realization of $\langle G,T \rangle $ vertices in $S$ must be located in a high location to be able to see all vertices in $S^\prime$. Moreover, according to the construction the sequence of vertices $o_i$, $o^{\prime}_i$, $s_i$, $r_i$, $s^\prime_i$ and $r^\prime_i$  lies on a chord-less cycle in $G$. While in any realization, internal faces are triangular and cycles with more that three vertices have chords (triangulation chords), the sequence of vertices $o_i$, $o^{\prime}_i$, $s_i$, $r_i$, $s^\prime_i$ and $r^\prime_i$ must lie on the outer boundary of the triangulation of any realization of the TIN. On the other hand, the visibility edges added in step $G\mhyph 1$ enforce the new vertices, except $o_i$'s and $o^\prime_i$'s, to be located as ridge points on the boundary of the triangulation to see other vertices in any realization of the TIN. However, the $o_i$'s and $o^\prime_i$'s must must be located as drain points of this boundary and have lower heights in such a way that their adjacent vertices $s_i$, $r_i$, $s^\prime_i$ and $r^\prime_i$ block their visibility.

Now assume that there is a realization for the instance $\mathcal{T}_\mathcal{A}$ of the \textit{StretchableTriangulation}. We can obtain a realization for the instance $\langle G,T \rangle$ of the visibility graphs of \textit{TIN}s as follows:

\begin{enumerate}
\item Scale and translate the planar realization of $\mathcal{T}_\mathcal{A}$ on the $XY$-plane to be surrounded by the circle $C$ with formula $x^2+y^2=1$.
\item Move the vertices $r_i$, $s_i$, $r^{\prime}_i$ and $s^{\prime}_i$ and the vertices in step $T\mhyph 2$, without interchanging their order on the boundary of the triangulation, to new locations on the $XY$-plane inside the circle with formula $x^2+y^2=4$ and outside the circle with formula $x^2+y^2=3$. This movement is done in such a way that each pair $(s_i,r_i)$ (resp. $(s^\prime_i,r^\prime_i)$) of vertices lie on different sides of the stretched pseudo-line $P_i$ with ultimately small distances from each other. This can be done by extending the stretched pseudo-lines in both sides to some points in the ring defined by the circles $x^2+y^2=4$ and $x^2+y^2=3$. Then, the points $r_i$, $s_i$, $r^{\prime}_i$ and $s^{\prime}_i$ and the vertices in step $T\mhyph 2$ are translated appropriately. Note that all intersection points of the stretched pseudo-lines are inside the circle $x^2+y^2=1$ and their extensions do not create new intersections.
\item Move the vertices $o_i$ and $o^\prime_i$ on the $XY$-plane to be on the intersection points (in both sides) of the underlying line of the stretched pseudo-line $P_i$ and the circle  $x^2+y^2=l$, for sufficiently large value of $l$. The value of $l$ is large enough so that all vertices of the stretched pseudo-line $P_i$ are inside the wedges defined by $\angle r_i o_i s_i $ and $\angle r^\prime_i o^\prime_i s^\prime_i$ and all other vertices lie outside.
\item Project the moved vertices in the second step on the truncated elliptic paraboloid (with downward convexity) which is truncated by the $XY$-plane with formula $x^2 + y^2=z +1$ and the moved vertices in the third step on the plane with formula $z=2$. The other vertices are left on the $XY$-plane and all the vertices in the second step will be located above the plane with formula $z=2$.
\end{enumerate}

In this procedure, the first 3 steps locate the projection of the points corresponding to the vertices of $\langle G,T \rangle$ on the plane and the next step locates these points in the space. Therefore, this realization supports the triangulation faces in $T$ as well as the visibility constraints in $G$. For the latter claim, the visibility graph of the flat part of the \textit{TIN} is the same as the triangulation which is not changed in our construction. The added vertices are also located on the second ($z=2$) and the third ($x^2 + y^2=z +1$) layers of the TIN match the visibility constrains. Therefore, this realization is valid for the $\langle G,T \rangle$ instance.

\begin{figure}[ht]
\centerline{\includegraphics[scale=0.27]{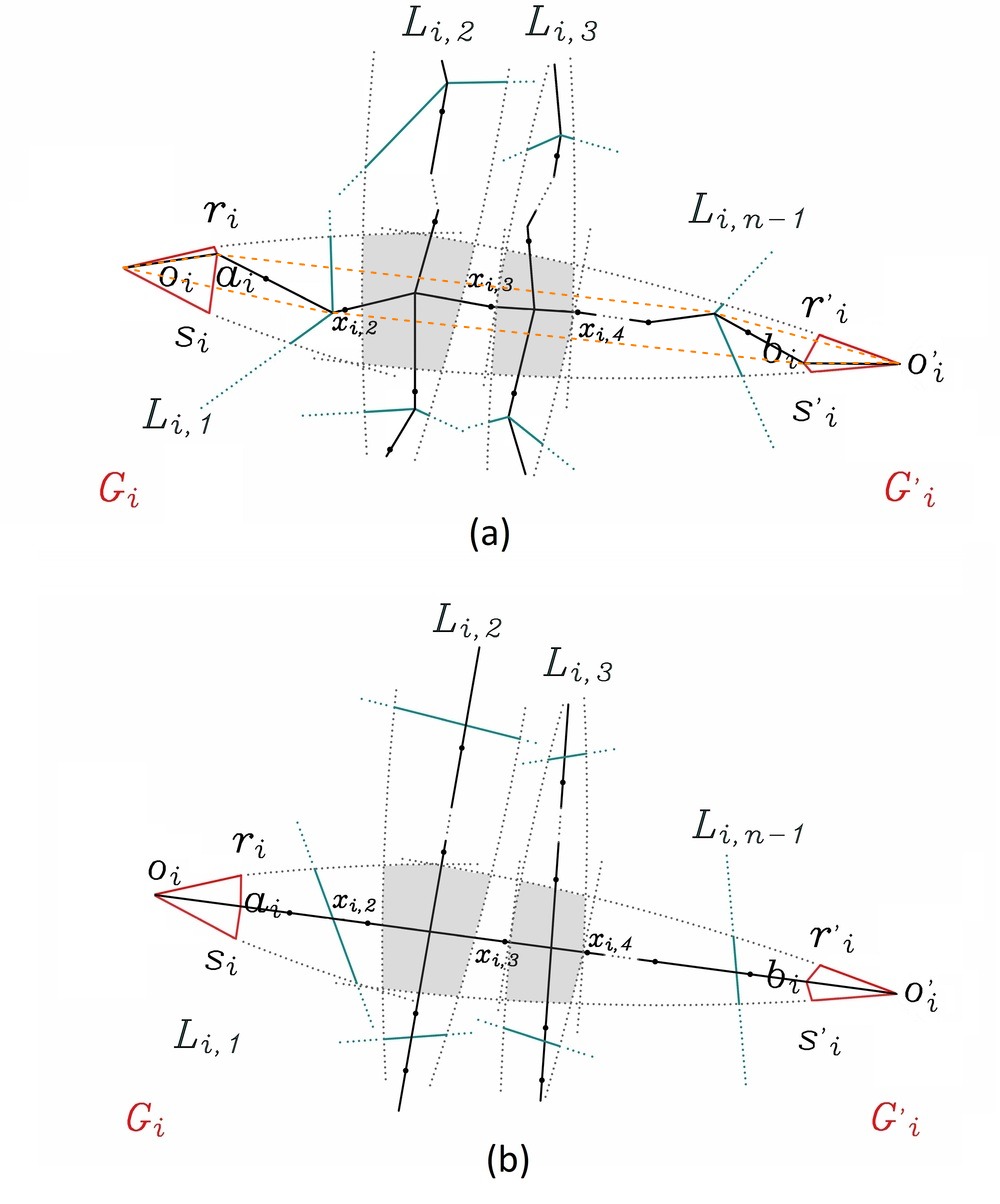}}
\caption{Stretching a pseudo-line $P_i$ in a realization of the $\langle G,T \rangle$ instance. \label{fig:6} }
\end{figure}

For the other side of the proof, assume that the instance of $\langle G,T \rangle$ has a \textit{TIN} realization. The projection of this realization on the $XY$-plane is a triangulation. Let's append a superscript star $(^*)$ to the name of each vertex in $G$, to denote its corresponding projected vertex on the plane. Let's $G_\mathcal{A}(P_i)$ be the set of vertices of $G_\mathcal{A}$ corresponding to the vertices of a pseudo-line $P_i$ (intersection points, splitting vertices, $o_i$ and $o^\prime_i$) in $\mathcal{A}$. All vertices of $G_\mathcal{A}(P_i)$ are visible from both $o_i$ and $o^\prime_i$. This implies that in this projected triangulation, the projection of these vertices must lie inside the intersection of the wedges defined by $\angle r^*_i o^*_i s^*_i $ and $\angle r^{*\prime}_i o^{*\prime}_i s^{*\prime}_i $. To prove this implication formally, we need some notations and then prove a claim which implies the correctness of this sentence. In a realization $R$ of $\langle G,T \rangle$, denote:
\begin{itemize}
\item the projection of $R$ on $XY$-plane as $R^*$
\item the set of realized points corresponding to the vertices of $G_\mathcal{A}(P_i)$ as $G^R_\mathcal{A}(P_i)$ and their projections on the $XY$-plane as $G^{R^*}_\mathcal{A}(P_i)$
\item the convex hull of $G^{R^*}_\mathcal{A}(P_i)$ as $C^{R^*}_\mathcal{A}(P_i)$ (see the orange convex hull in Figure~\ref{fig:6}) and its projection on $R$ as $C^R_\mathcal{A}(P_i)$
\item the realized part of $R$ corresponding to $I_{\mathcal{T}_\mathcal{A}}$ as $I_{\mathcal{T}_\mathcal{A}}(R)$ and its projection on $XY$-plane as $I_{\mathcal{T}_\mathcal{A}}(R^*)$ 
\end{itemize}

\begin{claim}
All vertices of $G^{R^*}_\mathcal{A}(P_i)$ are inside $C^{R^*}_\mathcal{A}(P_i)$ and all other projected points are outside it.
\end{claim}
\begin{claim_proof}
Clearly, all vertices of $G^{R^*}_\mathcal{A}(P_i)$ are inside their convex hull $C^{R^*}_\mathcal{A}(P_i)$. In addition, all vertices of this convex hull, lie inside the wedges defined by $\angle r^*_i o^*_i s^*_i $ and $\angle r^{*\prime}_i o^{*\prime}_i s^{*\prime}_i $, otherwise, they will be invisible to $o_i$ or $o^\prime_i$.

We prove by contradiction that all other projected vertices are outside $C^{R^*}_\mathcal{A}(P_i)$. For the sake of a contradiction, assume that there is a projected point ${p^*} \in R^*$ corresponding to a vertex $p \notin G_\mathcal{A}(P_i)$ that lies inside $C^{R^*}_\mathcal{A}(P_i)$. Denote the intersection of $C^{R^*}_\mathcal{A}(P_i)$ and $I_{\mathcal{T}_\mathcal{A}}(R^*)$ as $M^{R^*}_\mathcal{A}(P_i)$. Removing $M^{R^*}_\mathcal{A}(P_i)$ from $C^{R^*}_\mathcal{A}(P_i)$ split $C^{R^*}_\mathcal{A}(P_i)$ into two parts. Let $X^{R^*}_\mathcal{A}(P_i)$ be the part that includes $o^*_i$, and $Y^{R^*}_\mathcal{A}(P_i)$ the part that contains $o^{\prime*}_i$, and projections of $M^{R^*}_\mathcal{A}(P_i)$, $X^{R^*}_\mathcal{A}(P_i)$ and $Y^{R^*}_\mathcal{A}(P_i)$ on $R$ as $M^{R}_\mathcal{A}(P_i)$, $X^{R}_\mathcal{A}(P_i)$ and $Y^{R}_\mathcal{A}(P_i)$ respectively. $p^*$ lies inside one of $M^{R^*}_\mathcal{A}(P_i)$, $X^{R^*}_\mathcal{A}(P_i)$ or $Y^{R^*}_\mathcal{A}(P_i)$. 

\begin{itemize}
\item Assume that $p^*$ lies inside $M^{R^*}_\mathcal{A}(P_i)$ (hence, it is also located inside $I_{\mathcal{T}_\mathcal{A}}(R^*)$). $C^{R}_\mathcal{A}(P_i)$ which is included in $I_{\mathcal{T}_\mathcal{A}}(R)$ is a concave surface. In addition, all of the boundary points of $C^{R}_\mathcal{A}(P_i)$ are visible from $o_i$ and $o^\prime_i$ because each of them correspond to a vertex in $G_\mathcal{A}(P_i)$. It forces all points on the concave surface $C^{R}_\mathcal{A}(P_i)$ (including $p$) to be visible from $o_i$ and $o^\prime_i$ which contradicts with the invisibility of $o_i$ and $o^\prime_i$ to $p$ in $G_\mathcal{A}$. Therefore, $p^*$ can not be located in $M^{R^*}_\mathcal{A}(P_i)$. 
\item Assume that $p^*$ lies inside $X^{R^*}_\mathcal{A}(P_i)$. Therefore, some surfaces of $R^*$ blocks the visibility of $o_i$ and $p$. But, there is no vertex $q^*$ inside $X^{R^*}_\mathcal{A}(P_i)$ such that its projection on $R$ sees $o_i$. So, there must be a \textit{TIN} surface on $R$ whose triangular boundary is partially visible to $o_i$ and none of its boundary vertices is visible to $o_i$. The projection of such a triangulation boundary in $R^*$ will intersect the segment through $o^*_i$ and $a^*_i$ in a point which does not correspond to a vertex in $R^*$ and violates the monotonicity of the \textit{TIN}. Hence, $p^*$ cannot be located inside $X^{R^*}_\mathcal{A}(P_i)$. 
\item With a similar argument on the invisibility of $p$ and $o^\prime_i$, $p^*$ can not be located inside $Y^{R^*}_\mathcal{A}(P_i)$.
\end{itemize}

Therefore, $p^*$ can not be located inside $C^{R^*}_\mathcal{A}(P_i)$ and the claim is true.

\end{claim_proof}

For each pseudo-line $P_i$ in an arbitrary realization $R$ of $\langle G,T \rangle$, name an imaginary line through $o^*_i$ and $o^{*\prime}_i$ as $l_i$ (See Fig.~\ref{fig:6}-b) and do the following actions:

\begin{itemize}
\item For each vertex $p \in G_\mathcal{A}(P_i)$ which is the intersection vertex of pseudo lines $P_i$ and $P_j$ change the position of $p*$ to the intersection of $l_i$ and $l_j$.
\item Each vertex $p \in G_\mathcal{A}(P_i)$, which is not the intersection vertex of a pseudo-line $P_i$ with any other pseudo-line, lies between two intersection points $p^\prime$ and $p^{''}$ of the first type. For such an intersection point $p$, change the position of $p^*$ to an arbitrary point on $l_i$ between the new positions of points $p^\prime$ and $p^{''}$.
\end{itemize}

After these operations, all vertices of $G^{R^*}_\mathcal{A}(P_i)$ are moved to new positions on $l_i$. Because of the monotonicity of the pseudo-lines and $R$, the order of the vertices of $G^{R^*}_\mathcal{A}(P_i)$ along $l_i$ is the same as the order of their corresponding vertices on $P_i$. On the other hand, according to the proved claim, there is no vertex except vertices of $G^{R^*}_\mathcal{A}(P_i)$ inside the convex region $C^*_\mathcal{A}(P_i)$. Consequently, these displacements will not change the faces that any vertex $p$ belongs to.  Hence, these actions are done without changing the combinatorial structure of the triangulation $T$. Hence, the triangulation on these projected vertices is a realization for the instance $\mathcal{T}_\mathcal{A}$ of the \textit{StretchableTriangulation} problem. This completes the proof by stating that, if the instance  $\langle G,T \rangle$ of recognizing the visibility graphs of \textit{TIN}s is realizable, then there is a realization for the corresponding instance $\mathcal{T}_\mathcal{A}$ of \textit{StretchableTriangulation} problem and vice versa.
\end{proof}

Combining these results, we have the following theorem about the complexity of the recognizing problem:
\begin{theorem}
Recognizing visibility graphs of \textit{TIN}s is $\exists \Real\mhyph Complete$.
\end{theorem}
\begin{proof}
It is easy to show that recognizing visibility graphs of \textit{TIN}s belongs to $\exists \Real$ which is done by simply modeling an instance of this problem by an instance of satisfiability of a formula in $\exists \Real$.

On the other hand, Theorem~\ref{thm:2} states that \textit{Stretchability} is reducible to \textit{StretchableTriangulation} and  Theorem~\ref{thm:4} states that \textit{StretchableTriangulation} is reducible to recognizing visibility graphs of \textit{TIN}s, both in polynomial time. As  \textit{Stretchability} is $\exists \Real\mhyph Complete$, recognizing visibility graphs of \textit{TIN}s is $\exists \Real\mhyph Complete$ as well.
\end{proof}

\section{Conclusion}
In this paper, we showed that recognizing the visibility graphs of \textit{TIN}s problem is $\exists \Real\mhyph Complete$. We proved it by making a reduction from the \textit{stretchability} problem. 
As a possible future work, this result and technique may be useful in determining the computational complexity of other visibility graph recognition problems, particularly, recognizing the visibility graphs of monotone curves in two dimensions, whose computational complexity is still open.


\bibliographystyle{spmpsci}      
\bibliography{main}   

\pagebreak

\section*{Appendix}
\begin{algorithm}[ht]
\includegraphics[scale=0.90]{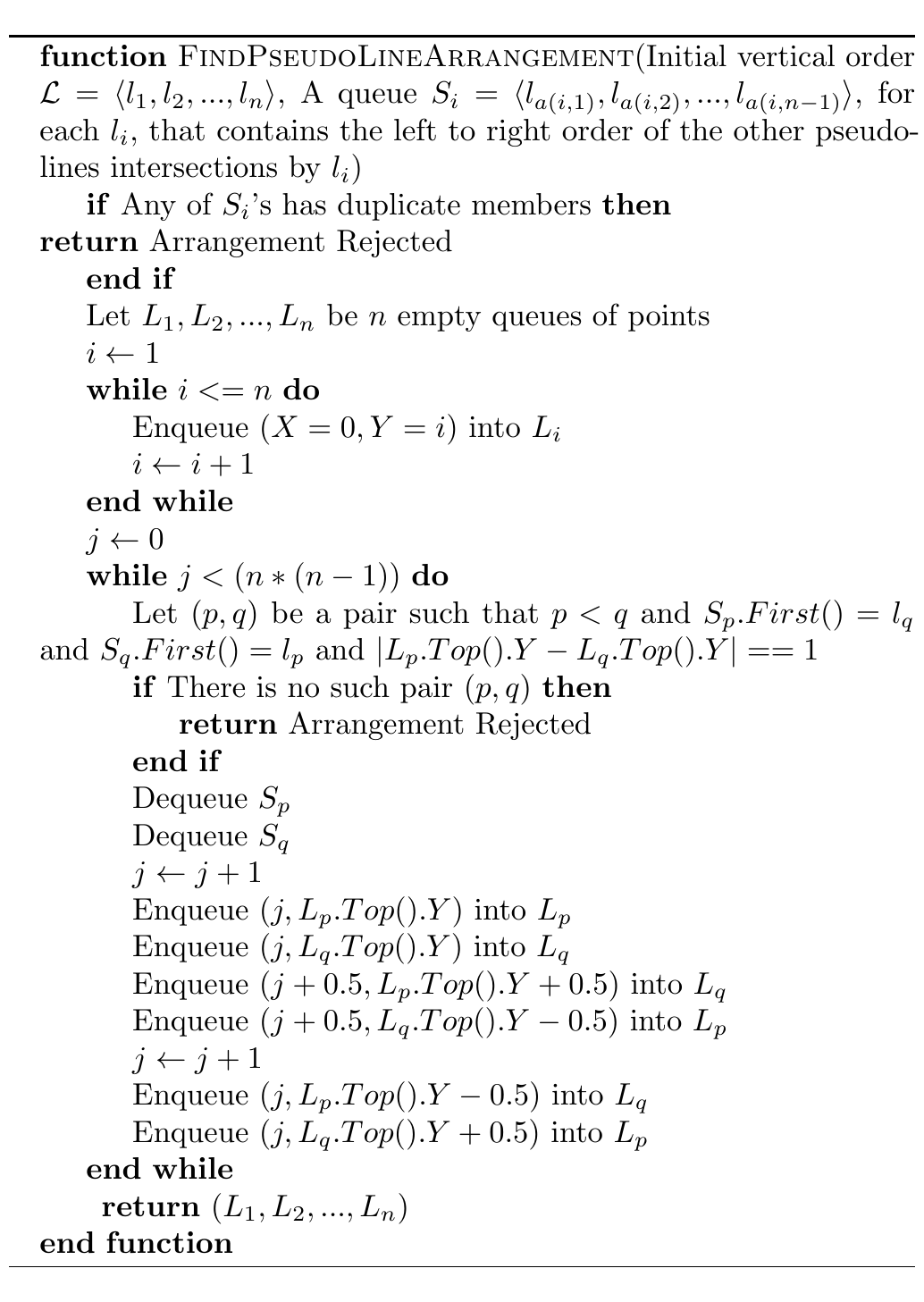}
\caption{Recognizing and Reconstruction algorithm for \textit{PseudoLineArrangement}\label{alg:1}}
\end{algorithm}

\end{document}